# DeepVRegulome: Deep learning framework for predicting the functional impact of short genomic variants on the human regulome

Pratik Dutta, [1]* Matthew Obusan[2], Rekha Sathian[1], Max Chao[1], Pallavi Surana[1], Nimisha Papineni[1], Yanrong Ji[3], Zhihan Zhou[4], Han Liu[4], Alisa Yurovsky[1], and Ramana V Davuluri[1]*
[1]Department of Biomedical Informatics, Stony Brook University, Stony Brook, NY, [2]Renaissance School of Medicine, Stony Brook University, Stony Brook, NY, [3]Division of Health and Biomedical Informatics, Department of Preventive Medicine, Northwestern University Feinberg School of Medicine, [4]Department of Computer Science, Northwestern University, Evanston, IL, USA.

## Abstract

**Background** Although whole-genome sequencing has revealed thousands of noncoding variants in cancer, characterizing their functional and clinical impacts remains challenging. Prioritizing mutations within gene regulatory regions is a critical bottleneck in genomics. Here, we present DeepVRegulome, a deep learning framework that systematically interprets the functional consequences of noncoding variants to identify clinically relevant mutations at a genome-scale.
**Methods** We developed DeepVRegulome, a computational framework integrating 464 fine-tuned DNABERT models (458 transcription factor, 4 histone mark, and 2 splice site models) trained on ENCODE and GENCODE datasets. The framework pairs these deep learning models with a suite of analytical tools: quantitative variant scoring via log-odds ratios to assess functional impact, attention-based motif analysis to identify disrupted sequence patterns, and survival analysis using Kaplan–Meier and Cox proportional hazards models to link high-impact variants with clinical outcomes. To ensure the framework accurately captures variant effects on baseline binding status, we benchmarked DeepVRegulome against an independent experimental assay of allele-specific transcription factor binding (SNP-SELEX) data and compared its performance to four established variant-effect predictors. Finally, we applied this pipeline to whole-genome sequencing data from 190 glioblastoma multiforme (GBM) patients across the TCGA and CPTAC-3 cohorts.
**Results** Across 65 transcription factors shared with the SNP-SELEX dataset, DeepVRegulome achieved a mean per-factor AUROC of 0.611. Following multiple-testing correction, 53 of 61 evaluable factors demonstrated a significant correlation between predicted variant-effect scores and experimental binding preferences. On a strict evaluation subset of 17 factors scored by all benchmarked methods, DeepVRegulome was significantly more accurate than DeepSEA and statistically indistinguishable from Enformer, Borzoi, and AlphaGenome—despite utilizing an input window 2–3 orders of magnitude smaller than these context-heavy models. Applying this validated framework to the glioblastoma (GBM) cohort yielded a comprehensive atlas of functional noncoding mutations, revealing thousands of high-impact somatic variants within the GBM regulome. Specifically, we prioritized 572 splice-disrupting and 9,837 transcription factor-binding site-altering mutations recurrent in over 10% of patients. Survival analysis demonstrated strong clinical relevance, directly linking 1,352 mutations and 563 disrupted regulatory regions to patient outcomes, which enabled patient stratification based on noncoding mutational signatures and highlighted novel prognostic markers in GBM.
**Conclusions** DeepVRegulome offers an interpretable and scalable solution to the challenge of deciphering noncoding genomic variation. Beyond mapping survival-associated mutations in glioblastoma, this framework establishes a generalizable, AI-driven paradigm for translating raw sequencing data into clinically relevant insights. By making the entire framework publicly available, we provide the genomics community with a powerful tool to drive discovery in functional regulation and precision oncology.

## 1 Introduction

Whole-genome sequencing (WGS)[1, 2] data across large cancer cohorts provide a comprehensive catalog of mutations, which include short variants, such as single nucleotide variants (SNVs), and small insertions and deletions (indels) [3-5]. While advances in variant calling have improved the detection of somatic mutations, interpreting their functional impact—particularly in noncoding regions—remains a major challenge[6, 7]. These noncoding mutations can affect splice sites, promoters, enhancers, and transcription factor-binding sites (TFBSs), thereby modulating gene expression and activity without altering protein structure[8, 9]. For example, mutations in TFBSs may disrupt TF binding, leading to aberrant gene expression[8], whereas alterations in splice sites may result in alternative splicing[10, 11], generating dysfunctional protein isoforms[12, 13]. The need for sophisticated computational models to interpret noncoding mutations is, therefore, paramount, as conventional annotation methods are not sufficient for these complex regulatory sequences[14, 15].

Recent advancements in deep learning methods have introduced models that yield promising results across biological applications, such as gene expression prediction[16], drug discovery[17], protein function analysis[18, 19], disease prognosis[20, 21], and noncoding variant effect prediction[9]. Specialized models for noncoding variant analysis include ARVIN[22], which identifies noncoding risk variants in promoters and enhancers; Basenji[23], which annotates mutations with chromatin accessibility changes; DeepBind[24], which predicts DNA/RNA-binding protein affinities; and DeepSEA[25], which assesses genomic variant effects on transcription factor binding, DNase I hypersensitive sites, and histone marks. These models address different aspects of gene regulation, offering powerful tools for variant impact analysis. Concurrently, large language models (LLMs), such as BigBird[26] and DNABERT[27,] leverage transformer architectures with k-mer tokenization to capture genomic sequence context, aiding in tasks such as masked language modeling (MLM). Extensions of this paradigm include epigenome-aware transformers (GeneBERT[28], Epigenomic-BERT[29]), which incorporate chromatin signals; long-context models (DNABERT-2[30], HyenaDNA[31]), which handle multispecies sequences; and cross-modal variants (DNAGPT[32], CD-GPT[33]), which jointly learn from genomic, RNA and protein data. While LLMs have also shown success in predicting coding variant effects (e.g., AlphaMissense[34], ESM1b-based models[35]) and other tools, such as FABIAN-variant[36,] focus on variant impact on TFBS via PWMs/TFFMs, there is a rising need for robust methods that specifically leverage LLMs to systematically integrate the genome-wide gene regulatory information and assess the impact of somatic short nucleotide variants across a comprehensive range of noncoding elements, as they capture both distal and proximal regulatory contexts.

Despite recent advances, few studies have implemented LLMs to systematically predict the effects of somatic mutations in noncoding regions, with existing tools often limited to small TFBS datasets and narrow genomic coverage. To address this methodological gap, we developed **DeepVRegulome,** a deep-learning framework integrating DNABERT-based[27] genomic foundation models, to systematically predict and characterize the functional impact of short genomic variants on both splice sites and a vast array of cis-regulatory elements, utilizing over 700 TFBS datasets from the Encyclopedia of DNA Elements (ENCODE) database[37]. We selected DNABERT over more recent models such as DNABERT-2 and Segment-NT because of its comparable performance on short sequences (less than 300 bp) relevant to these regulatory elements. Our framework fine-tunes DNABERT via two main sources of regulatory information: acceptor/donor splice sites from GENCODE and hundreds of ChIP-seq datasets from ENCODE for transcription factors and histone marks. This process yields highly accurate models designed to predict the functional impact of somatic mutations. DeepVRegulome further integrates quantitative variant scoring (log-odds ratios, score-change values), attention-based motif analysis for identifying disrupted motifs, and interfaces with established variant databases such as dbSNP[38] and ClinVar[39] for clinical context. We evaluated DeepVRegulome against four established variant-effect predictors, including AlphaGenome[40] and Borzoi[41]. For this benchmark, we used public SNP-SELEX data measuring the binding effects of 270 human transcription factors across 95,886 noncoding variants[42],

We applied DeepVRegulome to WGS data from 190 GBM tumors[43-47], an aggressive brain cancer with limited therapeutic options, to demonstrate its utility in a complex disease setting. The poor prognosis of glioblastoma and the known role of noncoding mutations in oncogenesis make it an ideal testbed for this approach[44, 48, 49]. Our analyses identified thousands of high-impact variants recurring across patients, including those disrupting critical regulatory elements linked to patient survival outcomes. The resulting variant signature robustly stratifies glioblastoma patients, highlighting clinically relevant noncoding mutations that are potentially valuable for therapeutic targeting[50, 51]. This study aims to provide a framework for uncovering novel, functionally significant noncoding mutations in all types of cancer, thereby offering new perspectives for precision oncology. In summary, **DeepVRegulome** represents a practical, efficient, and accessible contribution for interpreting noncoding somatic mutations, providing a powerful, generalizable tool for cancer genomics and precision oncology. Our openly accessible framework and interactive results dashboard facilitate future research into the regulatory genome, promoting broader adoption and extending its utility to other cancer types and genomic studies.

## 2 Datasets

To investigate the functional impacts of somatic genetic variants in noncoding regions associated with glioblastoma multiforme (GBM), we assembled three primary datasets:

a) *Whole-genome sequencing (WGS) data for glioblastoma patients*
b) *Splice Site Sequences*
c) *Transcription factor-factor binding site (TFBS) data*

### 2.1. Whole-genome sequencing data for glioblastoma patients:

We retrieved WGS data for glioblastoma patients from the Genomic Data Commons (GDC) Data Portal[52]. Specifically, sequencing data for 190 patients were obtained from the CPTAC-3 and TCGA-glioblastoma cohorts (**Supplementary Figure S2a**), which are extensively characterized and widely utilized in cancer genomics research. For each patient, we acquired two types of variant call format (VCF) files[53]: one generated via the CaVEMan variant caller[54] and another via Pindel[55].

2.2. **Splice Site and Transcription Factor-Binding Site Data:** For the splice site analysis, exon–intron junction sequences were sourced from the GENCODE project (version 40)[56]. Each junction was extended 45 nucleotides upstream and downstream to generate sequences of consistent length (90 bp) suitable for DNABERT input, resulting in hundreds of thousands of acceptor and donor windows. For the transcription factor-binding site (TFBS) analysis, binding site data were obtained from the Encyclopedia of DNA Elements (ENCODE) database[37], comprising data from 4228 ChIP-seq experiments for 667 TFs and 33 histone marks. The sequences were standardized to 301 bp, including 150 base pairs upstream and downstream of the binding site, to ensure a consistent input length for model training. Further details on dataset processing, characteristics, and summary statistics are provided in Section S2 of the supplementary material and Supplementary Figures S1 and S2.

# 3 Methods

## 3.1. DeepVRegulome Architecture overview

We fine-tuned DNABERT on two task-specific corpora—90-nucleotide sequences centered on GENCODE v41 exon–intron junctions for splice-site recognition and 301-nucleotide sequences centered on the ENCODE peak summits for 700 TFBSs. For splice sites, we trained separate acceptor and donor models, whereas for TFBSs, we trained one model per transcription factor, yielding 700 task-adapted classifiers. For each TFBS model, positive samples consisted of 301-bp sequences centered on ENCODE ChIP-seq peak summits. Negative samples were generated by randomly sampling length-matched sequences from the mappable human genome, excluding any window overlapping a positive peak region for the corresponding factor. Redundancy in the combined positive and negative pool was reduced using MMseqs2 clustering at 80% sequence identity, after which the retained sequences were downsampled to a 1:1 positive-to-negative ratio.

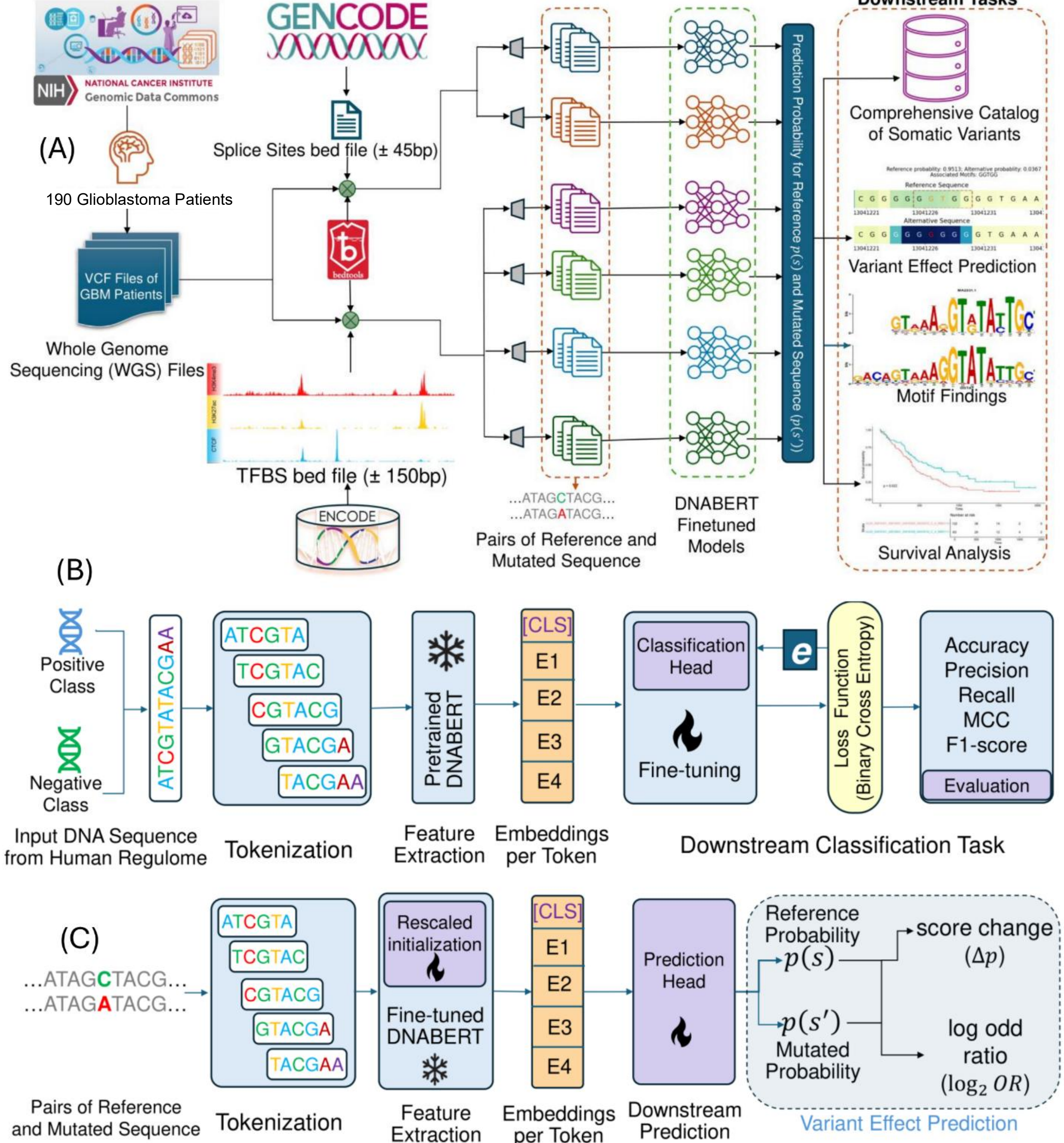

**Figure 1:** Architecture of **DeepVRegulome**. **(A)** Schematic of the computational framework: Starts with WGS and VCF data from 190 GBM patients (GDC). Splice site (GENCODE) and TFBS (ENCODE) annotations are intersected using BEDTools to define input regions, from which reference and mutated sequence pairs are extracted. High-impact variants are then cataloged for downstream analysis, including motif disruption and clinical annotation. **(B)** Pretrained DNABERT model is fine-tuned for regulatory prediction tasks using labeled data from donor/acceptor splice sites and ENCODE ChIP-seq peak regions. Fine-tuning is performed in two stages: transfer learning from the pretrained model (frozen or partially trainable) and training from scratch for each regulatory class. **(C)** Fine-tuned models are used to score somatic variants in GBM patient genomes by computing prediction probabilities on wild type (p(S)) and mutated sequences (p(S')). Variant disruption scores are derived via log-odds ratio and attention-based score change.

The balanced dataset was split randomly into training, validation, and test sets at an 80:10:10 ratio. All reported classification metrics were computed on the held-out test set. The same construction was applied uniformly across all TFBS models.. A detailed description of the fine-tuning data collection is provided in the "Data Acquisition" section of the supplementary material. The detailed parameters and procedures for the DNABERT fine-tuning process, including k-mer tokenization, encoder architecture, classification head design, training regimen, hyperparameter optimization, and performance evaluation metrics, are provided in **Supplementary file S1**.

## 3.2. Variant impact prediction and scoring

Once we finalized the selection of the best finetuned models, every substitution or indel located within the corresponding noncoding region was evaluated. To achieve this goal, a systematic data preprocessing workflow was established to prepare the datasets for input into these fine-tuned DNABERT models. The process began by identifying variants located within splice sites and TFBS regions via the bedtools[57] intersect module, effectively pinpointing mutations within these critical noncoding regions. For every intersecting event, we constructed a matched sequence pair: the reference (wild-type) sequence $(s)$ and the corresponding mutated sequence $(s')$, bearing the alternative allele. These sequence pairs constituted the inputs for the fine-tuned DNABERT models in all downstream scoring analyses.

A variant was classified as functionally disruptive if the model's posterior probability (p(s)) of the wild type was greater than 0.5 and the probability of the mutated sequence (p(s')) was less than 0.5. Disruption magnitude is quantified via two complementary statistics: score change[24]: $S = \Delta p = \big(p(s') - p(s)\big)\max(p(s'), p(s))$, where the max term is added to amplify the strong effects of genetic variants; and the log odds ratio[25] $\log_2 OR = \log_2 \frac{p(s)}{1-p(s)} - \log_2 \frac{p(s')}{1-p(s')}$, which reflects the association between events "being classified as positive" and "having the particular genetic variant"

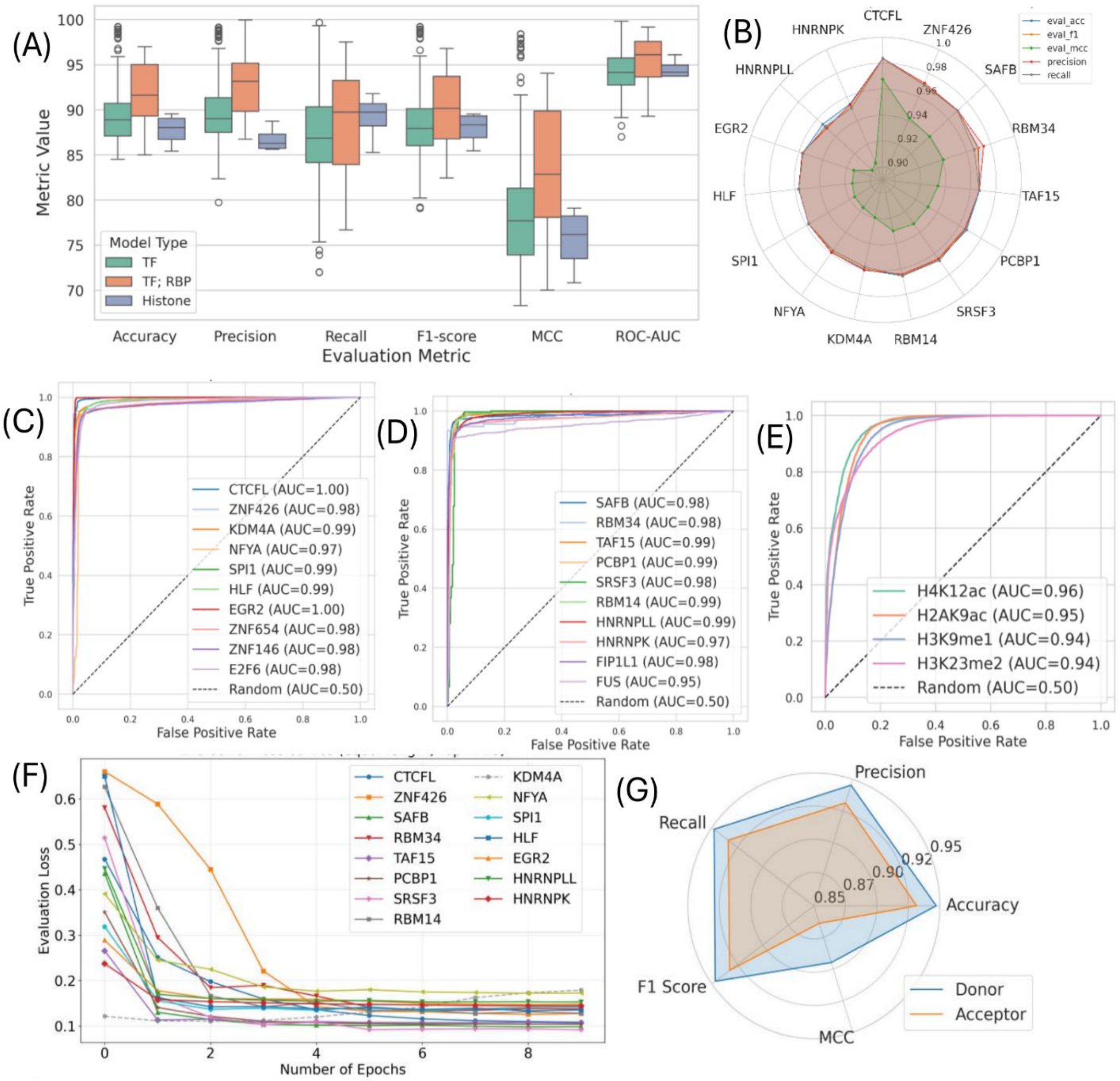

Figure 2: **Performance of DeepVRegulome models in predicting diverse regulatory elements.** (a) Boxplots comparing six evaluation metrics across fine-tuned models for transcription factors (TFs; n = 421), RNA-binding proteins (RBPs; n=37), and histone marks (n = 4). The results highlight consistently high performance for TF and RBP models and more variable performance for histone mark models. (b)Radar plot showing multimetric performance of the top 15 TFBS models, selected based on evaluation accuracy, revealing uniformly high recall and MCC across top performers. (c,d) ROC curves for the top 15 TFBS models (c) and top 15 RBP models (d) demonstrate excellent discriminative ability, with AUC values ≥ 0.98 and ≥ 0.95, respectively. (e) ROC curves for all four histone marker models which having evaluation accuracy greater than 85%, indicating slightly lower but still robust classification performance. (f) Evaluation loss trajectories across top 15 TFBS models, showing steady convergence and stable generalization behavior. (g) Comparative radar plot between donor and acceptor splice site classifiers across five evaluation metrics, emphasizing trade-offs between precision and recall. Collectively, these results highlight the robustness and generalizability of DeepVRegulome across diverse classes of regulatory elements.

(**Figure 1C**). A larger log odds ratio (>0) indicates that the variant results in more impactful functional disruption.

## 3.3. Output and Application

This pipeline converts raw WGS data into quantitative estimates of regulatory disruption. The resulting catalog comprises per-variant probability shifts, $\Delta p$ scores and $\log_2$ odds ratios for every splice site and TFBS mutation observed across our glioblastoma cohort, enabling direct prioritization of noncoding variants for downstream survival analysis, motif interrogation and experimental validation. Applying DeepVRegulome thus revealed candidate variants with putative effects on gene regulation and splicing in glioblastoma, offering new insights into the contribution of noncoding variation to tumor biology.

# 4 Results

DeepVRegulome is an integrated computational framework composed of four distinct modules designed to systematically predict and interpret the functional impact of noncoding variants (Figure 1).

## 4.1. Regulatory Region Prediction Module:

To perform accurate in silico mutagenesis experiments on short DNA sequences, highly accurate classification models for predicting functional gene regulatory DNA sequences are needed. We developed DNABERT-based fine-tuned models for predicting regulatory regions of two different sizes, (i) ±45 bp regions flanking splice sites and (ii) ±150 bp regions flanking ChIP-Seq (for TFBS and histone marks) or RNA Bind-n-Seq peak regions, covering 23% of the human genome.

The splice site models demonstrated high predictive accuracy across multiple metrics (accuracy: 94–96%; F1 score: 0.94–0.96; MCC: 0.89–0.92; Figure 2f). This performance underscores the ability of DNABERT to learn the complex sequence features surrounding canonical splice junctions. For ENCODE regulatory elements, we fine-tuned models on ChIP-seq peak regions for 667 TFs and 33 histone marks; we also designated 59 of the TFs as RNA-binding proteins (RBPs) on the basis of their known functions[58]. We included only those fine-tuned models that surpassed 85% accuracy. While 421 TFs (**~63%** of all the evaluated) and 37 RBP models achieved this accuracy, only 4 histone marks (H4K12ac, H2AK9ac, H3K9me1, and H3K23me2) were included in the final set of fine-tuned models. This demonstrates the strong ability of DNABERT fine-tuned models to learn sequence-specific patterns in TF target regions. In contrast, performance was more variable for histone marks, suggesting that their positioning may be governed by factors beyond local sequence content that are not fully captured by the current models.

The top-performing TFBS model (CTCFL[59]) achieved close to perfect accuracy (0.9838), followed by ZNF426[60] (0.9709), SAFB[61] (0.9702), RBM34[62] (0.97) and TAF15[63] (0.9674). These results confirm the robustness and specificity of DNABERT models across a wide variety of TFs, including both canonical regulators (e.g., SPI1, E2F6, MITF, USF1) and RNA-binding proteins (SRSF3, RBM14, PCBP1). As shown in the boxplots (Figure 2a), the TF models and RNA-binding proteins consistently demonstrated high predictive performance across all the metrics, whereas the histone marker models, although fewer in number, also achieved robust classification with slightly broader variance. Receiver operating characteristic (ROC) curves for the top 15 TFBS models (Figure 2c) revealed exceptional discriminative power (AUC > 0.97), further supported by radar plots (Figure 2b), indicating uniformly high recall and MCC values across the top-performing models. The evaluation of the loss trajectories confirmed stable convergence and generalization across training epochs (Figure 2f). In parallel, all four histone marker models exhibited strong ROC performance (AUC range: 0.94–0.96; Figure 2e), demonstrating that the fine-tuned models in this module effectively capture regulatory signals at base-pair resolution for both transcription factors and chromatin features.

These results establish the robustness and discriminative capacity of DeepVRegulome for modeling TF-specific DNA binding signatures, even under the variability and noise inherent to ChIP-seq data. The consistent performance across both broadly expressed and tissue-specific TFs demonstrates the framework's suitability for downstream variant impact prediction. Collectively, the high-fidelity models for both splice sites and TFBSs form the predictive backbone of DeepVRegulome, enabling reliable assessment of noncoding variant effects in disease-relevant contexts.

## 4.2. In silico Mutagenesis Module:

The main objective of this module is to predict the effects of short variants on a regulatory sequence, such as a TF or histone-bound region, splice sites, or RBP binding sites. First, the mutation information from VCF files is integrated with ENCODE regulatory regions and splice sites to find all the mutated regulatory regions in the genome. For a given regulatory sequence with mutations, we systematically change the nucleotides within a sequence and use a computational model to predict the resulting change in the sequence's function or property on the basis of predictive probability or score changes from the corresponding fine-tuned model compared with the reference sequence.

We applied our fine-tuned models to predict the functional impact of short variants from 190 glioblastoma patients in the TCGA and CPTAC-3 cohorts. We analyzed variants within defined splice site regions (±45 bp around junctions) and cis-regulatory regions (±150 bp around ChIP-seq peak summits for TFBSs and histone marks). The functional impact was subsequently quantified via a log-odds ratio and a prediction score change between reference and mutated sequences. The analysis revealed a high burden of somatic mutations within noncoding regulatory elements, particularly in TFBS regions. Within these sites, we identified 932,133 SNVs mapped across 358,414 distinct regions and over 2.13 million indels mapped across 533,433 regions. DeepVRegulome predicted 35,867 indels and 19,709 SNVs to be functionally impactful in these regions, of which 8,598 indels and 1,087 SNVs were recurrent in ≥10% of patients. The same analysis was performed for splice sites (acceptor and donor), histone markers, and transcription factor-binding sites. A detailed summary of all identified variants, with further categorization of those predicted as functional and those recurring in more than 10% of patients, is provided in Table 1 and Supplementary Section S4.

Finally, we exclusively focused our biological and clinical analyses on the predicted functional variants observed in ≥10% of glioblastoma patients—a threshold chosen to balance the

**Table 1:** Summary of total number of variants in the regulatory regions and predicted functional variants and affected regulatory regions across splice site and ChIP-seq–based fine-tuned DNABERT models.

| Fine-tuned models | Subtype | Variant Type | Number of | | Predicted variants and regulatory regions | |
|---|---|---|---|---|---|---|
| | | | Variants | Regions | Number of Variants (present in >10% of the samples) | Number of Regions (present in >10% of the samples) |
| Splice Sites | Acceptor | SNVs | 19,968 | 14,743 | 299 **(1)** | 265 **(3)** |
| | | Indels | 34,718 | 19,897 | 1,822 **(437)** | 1,375 **(408)** |
| | Donor | SNVs | 23,656 | 15,849 | 673 **(4)** | 545 **(4)** |
| | | Indels | 20,171 | 13,491 | 504 **(130)** | 423 **(122)** |
| ChIP-seq Models | Histone Markers | SNVs | 127,297 | 57,013 | 387 **(2)** | 311 **(5)** |
| | | Indels | 254,566 | 87,949 | 536 **(150)** | 373 **(129)** |
| | TFs | SNVs | 932,133 | 358,414 | 19709 **(1087)** | 16566 **(322)** |
| | | Indels | 2,130,274 | 533,433 | 35867 **(8598)** | 30411 **(7,297)** |

## 4.3. Benchmarking with SNP-SELEX data

The preceding analyses establish that DeepVRegulome classifies bound and unbound regulatory sequences accurately and recovers known binding motifs. A separate question is whether the model's output probabilities track the effect of a variant on binding, rather than binding status alone. To test this directly, we evaluated DeepVRegulome against an independent experimental measurement of allele-specific transcription-factor binding: the SNP-SELEX assay of GSE118725[42], which quantifies preferential binding between reference and alternate alleles in vitro. For each variant we scored both alleles with the relevant factor's model and used the signed difference in predicted probability as the variant-effect score. Because the models were trained only on ENCODE ChIP-seq peaks and never exposed to SNP-SELEX data, this constitutes an out-of-distribution test against an orthogonal assay. Across the 65 transcription factors shared between our model set and SNP-SELEX, DeepVRegulome achieved a mean per-factor AUROC of 0.611 (median 0.601) over the 61 factors with sufficient positive variants (n_positive ≥ 5) to compute the statistic. After Benjamini-Hochberg correction, 53 of the 61 evaluated factors showed a significant correlation between predicted variant-effect scores and experimental binding preference ($q < 0.05$), confirming that the model's scores carry information about variant effects on binding beyond chance (Supplementary Table S11).

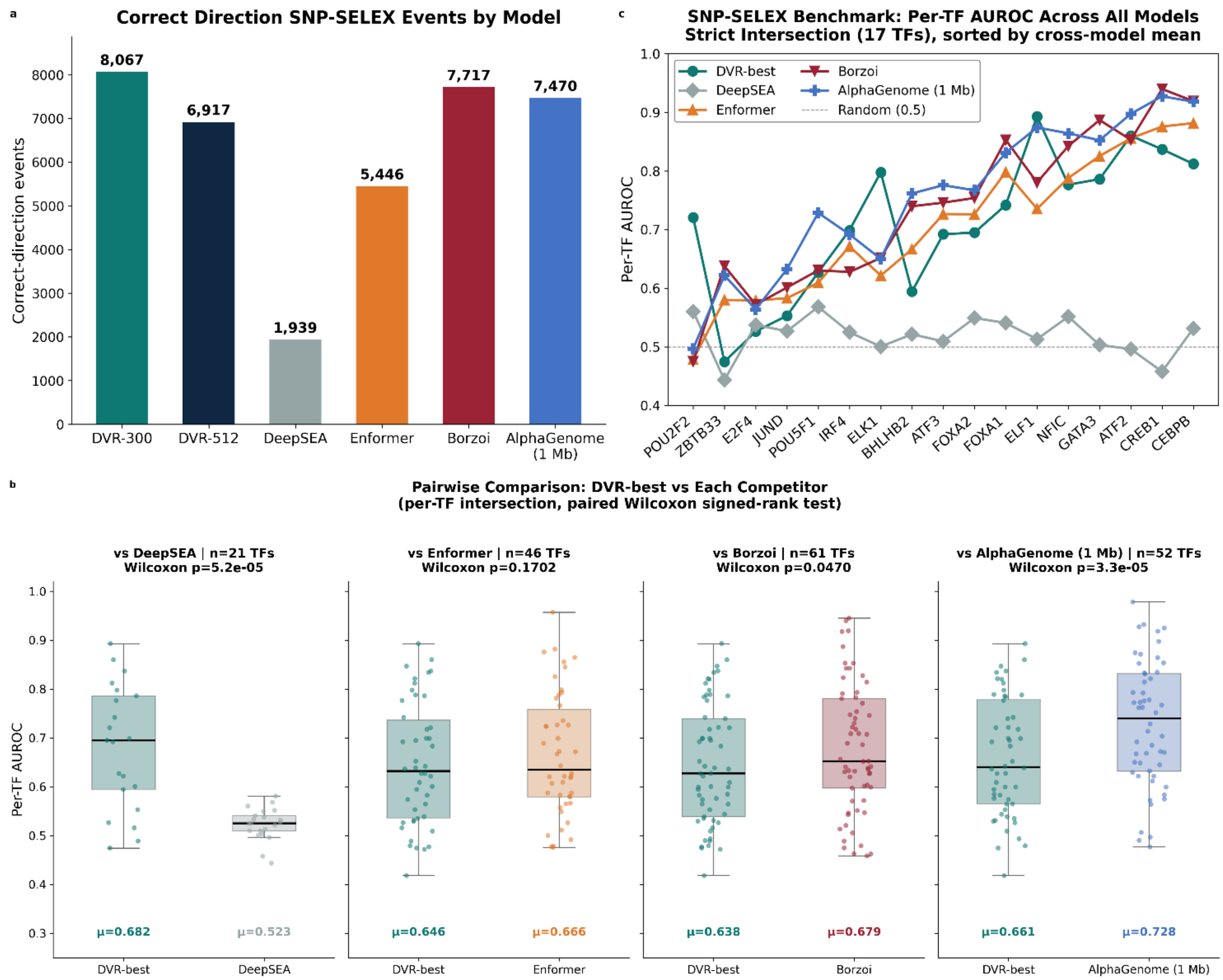

**Figure 3.** DeepVRegulome benchmarking against competitor models on the SNP-SELEX variant-effect prediction task. (a) Number of correct-direction predictions per model across the 10,513 experimentally confirmed allelic imbalance events (|PBS| > 1.0, $p < 0.05$) spanning the 65-TF DVR-evaluable universe. (b) Pairwise comparison of DVR-best (per-factor maximum AUROC across the 301 bp and 512 bp models) against each competitor on their shared evaluable transcription factors, using a paired Wilcoxon signed-rank test. Box plots show the interquartile range with individual TF values overlaid; sample sizes (n) and mean AUROC (μ) are reported for each group. (c) Per-TF AUROC across all six methods on the strict 17-transcription-factor intersection scored by every method, sorted by cross-model mean. On this matched subset, DVR-best performance tracks closely with the long-context foundation models across most factors, with no single method dominating uniformly. Dashed line indicates random performance (AUROC = 0.5).

Beyond per-factor discrimination, we assessed each method's ability to predict the direction of allelic effects across the full SNP-SELEX event space. Of the 10,513 experimentally confirmed allelic imbalance events ($|PBS| > 1.0$, $p < 0.05$), DVR-300 correctly predicted the direction for 8,067, the highest of any method tested (Figure 3A). Borzoi[41], the closest competitor in TF coverage, correctly predicted 7,717 events, followed by AlphaGenome[40] (7,470), DVR-512 (6,917), Enformer[16] (5,446), and DeepSEA[25] (1,939). DeepVRegulome thus matches or exceeds every competitor on directional concordance with the experimental ground truth despite its smaller input window.

To assess whether DeepVRegulome's input window limits variant-effect prediction, we re-fine-tuned all 65 evaluable factor models at a wider 512 bp window and re-evaluated them on the same SNP-SELEX dataset. Although the wider context improved classification accuracy on held-out development data, variant-effect AUROC showed a modest trend toward improvement but did not differ significantly from the 301 bp models (mean AUROC 0.626 vs 0.611; paired Wilcoxon $p = 0.076$). Of 59 factors evaluated at both window sizes, 37 favored the 512 bp model and 22 favored the 301 bp model. These results indicate that while wider context provides a small benefit, the sequence grammar governing TF binding site disruption is largely captured within a few hundred base pairs, consistent with the observation that TF ChIP-seq and chromatin accessibility predictions can be made accurately from approximately 500 to 2,000 bp of surrounding sequence. Because the difference between window sizes was modest and did not reach statistical significance, we defined

"DVR-best" as the per-factor maximum AUROC across the 301 bp and 512 bp models (mean AUROC 0.638) and used this composite score for all subsequent competitor comparisons.

As the most conservative head-to-head comparison, we identified the 17 transcription factors scored by every method (DVR-best, DeepSEA, Enformer, Borzoi, and AlphaGenome) and compared per-factor AUROC on this identical set (Figure 3C). On these 17 factors, DVR-best was significantly more accurate than DeepSEA (paired Wilcoxon $p = 3.1 \times 10^{-5}$) and was statistically indistinguishable from Enformer, Borzoi, and AlphaGenome ($p = 0.353$, 0.329, and 0.051 respectively). Figure 3C shows per-factor AUROC across all five methods on this matched set; no single method dominates every factor, and the long-context models pull ahead mainly on the harder half of the distribution (lower cross-model mean AUROC). DeepVRegulome reaches this performance using an input window of 301 to 512 bp, two to three orders of magnitude smaller than the 197 kbp (Enformer), 524 kbp (Borzoi), or 1 Mb (AlphaGenome) context of the competing models.

Expanding the comparison to each method's full set of evaluable factors reveals a gradient that tracks input context length. On the 21 factors shared with DeepSEA, DVR-best significantly outperformed it (mean AUROC 0.682 vs 0.523; $p = 5.2 \times 10^{-5}$). On 46 shared factors with Enformer, the two methods were not significantly different (0.646 vs 0.666; $p = 0.17$). On 61 shared factors with Borzoi, Borzoi held a modest edge (0.638 vs 0.679; $p = 0.047$), and on 52 shared factors with AlphaGenome, AlphaGenome led more decisively (0.661 vs 0.728; $p = 3.3 \times 10^{-5}$; Figure 3B). Within AlphaGenome itself, the 1 Mb configuration significantly outperformed its 16 kbp configuration on this dataset ($p = 0.003$), providing direct evidence that longer genomic context improves variant-effect prediction for a subset of factors. A threshold analysis further supports this interpretation: the AUROC gap between DVR-best and AlphaGenome narrows from 0.067 at a minimum of 5 positive instances per factor to 0.028 at a minimum of 200, where the difference is no longer significant ($p = 0.099$; Supplementary Figure S4). The residual advantage of AlphaGenome thus concentrates among factors with few positive examples, consistent with a context-dependent benefit for specific factors rather than a systematic architectural deficit. Full per-factor results are provided in Supplementary Table S11.

## 4.4. Motif Visualization and Interpretation Module:

To facilitate model interpretability, DeepVRegulome includes a module for the post hoc analysis of fine-tuned models. This module first derives candidate motifs from learned models, which are then subjected to rigorous biological validation. Candidate motifs are identified via two primary strategies: statistical enrichment and attention-based analysis. First, we performed k-mer enrichment analysis between positively and negatively labeled sequences via a hypergeometric test, identifying cis-regulatory elements significantly overrepresented in ChIP-seq-bound regions. In parallel, attention weight heatmaps from the DNABERT layers pinpoint salient nucleotides, from which sequence motifs crucial for model prediction can be extracted. This dual approach enables interpretability at both the sequence and variant levels, enhancing the biological relevance and explainability of DeepVRegulome predictions.

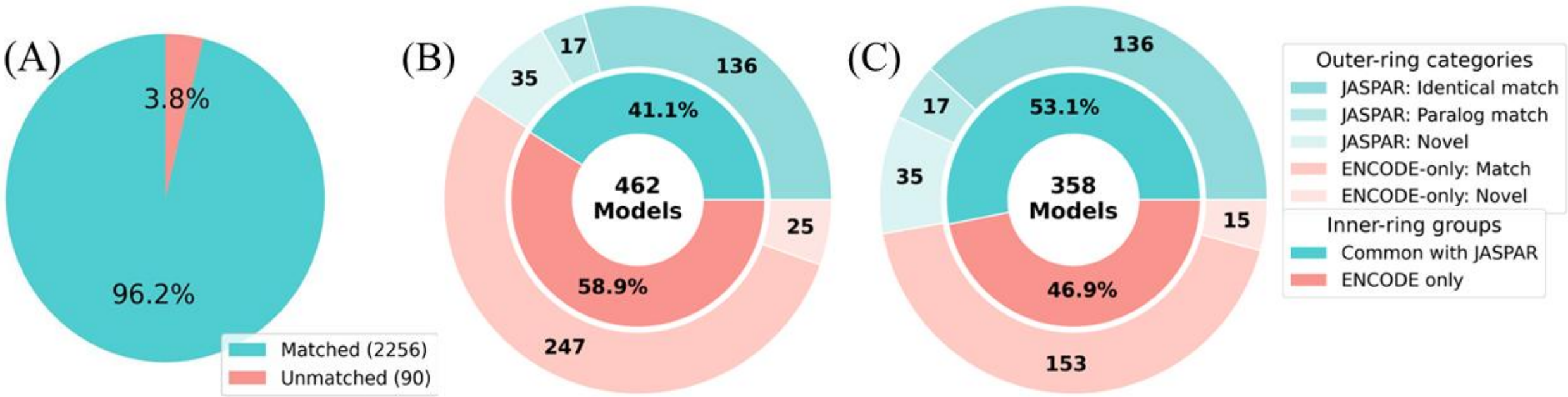

**Figure 4**: **Large-scale validation and characterization of learned regulatory motifs. by DeepVRegulome.** (A) Comprehensive coverage of the known regulatory sequence space. The 402 models with significant JASPAR similarity collectively recapitulate 96.2% (2,256 of 2,346) of the curated matrices in the JASPAR database. (B) Hierarchical breakdown of 462 high-confidence models following comparison to the JASPAR 2024 database (Tomtom, $q \leq 0.05$). The inner ring shows the initial split between models for TFs common to both ENCODE and JASPAR (n=190) and those unique to the ENCODE-derived set (n=272). The outer ring details the validation results for each group, classifying motifs as identical matches, paralog matches, or novel. Percentages indicate the proportion within each of the two primary groups. (C) The same breakdown restricted to the 358 bona-fide sequence-specific DNA-binding TFs per the Lambert et al. 2018 census. All 190 common-with-JASPAR models are DNA-binding TFs by this classification, so the inner and outer rings of the common-with-JASPAR half are unchanged from panel B.

To assess the biological accuracy of the framework, we rigorously evaluated the motifs learned by our 462 high-confidence TF, RBP and Histone models against the JASPAR 2024 database. Of these 462 models, 358 (77%) correspond to bona-fide sequence-specific DNA-binding domain (DBD) transcription factors per the Lambert et al. 2018[64] human TF census; the remaining 104 comprise cofactors and chromatin-associated proteins (65 models, 14%) and factors not present in the Lambert census including histone marks and RNA-binding proteins (39 models, 8%).

To identify if our models could recapitulate their associated known DNA motifs, we selected for an overlap of our high performance DBD TFs ENCODE-derived catalog (358) matched against JASPAR Database, of which 190 transcription factors were found present in both via an HGNC symbol match (**Figure 4C**). Within this shared set, 136 (~72%) of the models learned motifs identical to their designated JASPAR counterparts (Supplementary Figure-S3). An additional 17 models (~10%) matched the known motif of a closely related paralog from the same TF family. This is exemplified by the model for SP1, which learned a motif that strongly matched the JASPAR profile for its paralog, Klf6-7-like ($q$=2.89e-07). This result correctly captures the conserved binding preference of the large Sp/KLF family, whose members share a canonical three-zinc-finger domain that recognizes GC-rich promoter elements. In total, 82% of this direct validation cohort (153 of 190) identified a biologically plausible binding preference. The remaining 35 models (~18%) represent potential discoveries of novel or alternative binding specificities for established TFs.

Next, we characterized the remaining 168 (of the original 358) DBD TFs models that lacked a direct name match in the JASPAR database. The vast majority of these models, 153 models (91%), still produced motifs with significant similarity to a JASPAR motif annotated to another TF, leaving 15 completely unmatched PWMs. A hierarchical breakdown of the validation results for all 462 models is provided in Figure 4B.

Finally, an analysis of the cofactor TFs (65) allows us to understand where their known associated DNA recruitment areas correspond to other known DBD TFs. 60 of the 65 TFs in this subset identified 2014 unique JASPAR motifs that correspond to other DBD TFs in the region, with many being shared across the cofactors TF models (1322 of 65.6% of identified unique DBD motifs are shared across at least 10 of the cofactors). Of these, some of the most frequently recovered targets are CTCF (60/60), KLF/SP family motifs (60/60), GATA1::TAL1 (58/60), and other sequence-specific DNA-binding factors that are known to recruit chromatin modifiers and coactivators to their target sites, showing the power of these models to learn the sequential biology associated with recruiting cofactor TFs (results summarized in Supplementary Table 10).

Novelly, the comparison revealed that 402 of our 462 models (87%) learned motifs with significant similarity to a known JASPAR profile (Tomtom, $q \leq 0.05$). In fact, when matching all identified motifs in our 462 model set, 402 models identify 2256 (out of 2346) unique binding motifs found in JASPAR, signifying the binding enrichment and associations that are often found near these binding sites. For example, a G>C variant within an ETS1 motif upstream of the *GATAD2B* gene was shown to disrupt the core CCCGA sequence, which directly corresponded to an attenuation of DNABERT's attention scores at that site (Fig. 5B). These results confirm that *DeepVRegulome* not only achieves high performance in classification but also captures biologically interpretable features aligned with established regulatory signatures. This interpretability module thus bridges deep learning predictions with biological knowledge, allowing researchers to assess not only whether a variant is impactful but also why—by tracing predictions back to recognizable motif regulatory grammar.

## 4.5. Survival analysis and patient stratification module:

To assess the clinical relevance of the variants identified by DeepVRegulome, we performed survival analysis and found that predicted functional disruptions in both TFBSs and splice sites were significantly associated with patient outcomes. For TF binding sites, we found that the vast majority of these clinically relevant variants represent novel findings. An evaluation against the ClinVar database revealed that only 13 significant indel variants from our top TFBS models had prior clinical entry, with no SNVs having a record. The few variants that were found in TFs with established roles in cancer progression. These include key regulators such as SMAD4 (which plays a role in pathways such as TGF-β signaling)[65] and FOXM1, which drives cell proliferation, particularly in oncogenesis. The discovery of these functionally impactful variants, despite their absence from clinical databases, highlights a central point: there is a large, clinically uncharacterized landscape of noncoding mutations that DeepVRegulome can help prioritize.

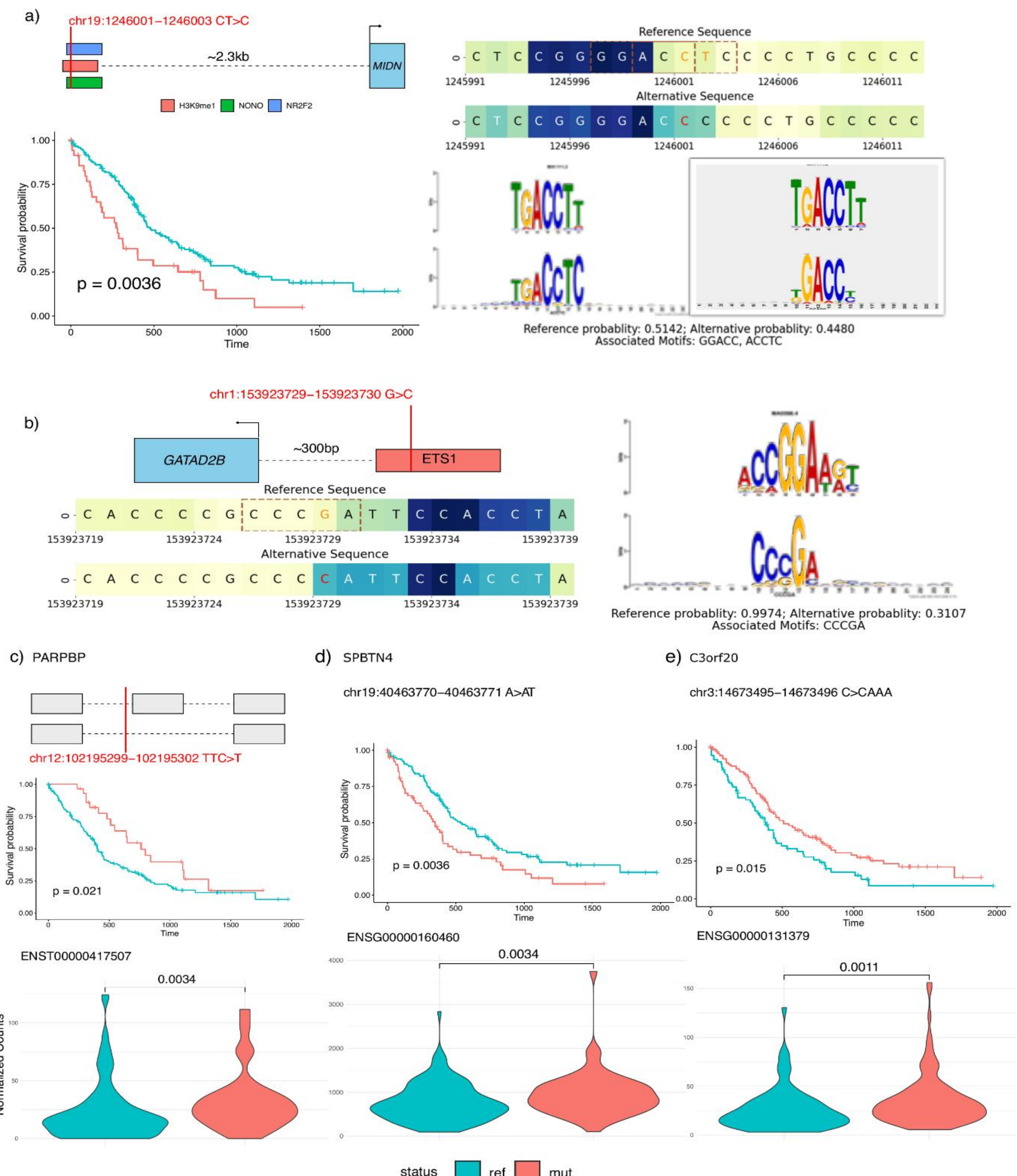

Figure 5: **Integrative analysis of high-impact noncoding variants and their association with clinical outcomes in glioblastoma.** (A) A CT>C substitution in a predicted NR2F2 binding site located ~2.3 kb upstream of the MIDN gene. The mutation is predicted to be disruptive, as shown by the attention score heatmaps (right). Kaplan–Meier analysis (left) shows that patients harboring this variant have significantly worse overall survival ($p = 0.0036$). (B) A G>C substitution in a predicted ETS1 binding site located ~300 bp upstream of the GATAD2B gene. The model predicts a strong disruption of the core motif, with binding probability dropping from 0.99 to 0.31. (C) A TTC>T variant in a splice site region is associated with worse patient survival ($p = 0.021$) and is linked to significantly increased expression of a transcript missing an exon downstream of the disruption, ENST00000417507 ($p = 0.0034$). (D) An A>AT insertion is associated with worse survival ($p = 0.0036$) and a significant increase in the expression of the gene SPBTN4 ($p = 0.0068$). (E) A C>CAAA insertion is associated with worse survival ($p = 0.015$) and a significant increase in the expression of the gene C3orf20 ($p = 0.0021$).

Disruptions in TFBSs also strongly correlated with clinical outcomes. For example, a substitution in a PNOX1 TFBS within the gene *TLK2* was associated with significantly worse survival (HR = 2.3, $p = 2.7 \times 10^{-3}$). TLK2 has recently been identified as a potential therapeutic target, and inhibitors of this gene have shown efficacy in several cancer cell lines[66]. Similarly, a deletion in a RELB TFBS within the *ZDHHC8* gene, which is correlated with worse survival when highly expressed, also predicted poorer outcomes (HR = 2.1, $p = 1.2 \times 10^{-4}$)[67]. Conversely, an insertion in an SP1, a TFBS upstream of a transcript of the gene *PRKCE* was associated with better survival (HR = 0.46, $p = 2.5 \times 10^{-2)}$. Protein kinase C and its isozymes have been intensely studied as therapeutic targets in glioblastoma[68-70]; however, very few works in the literature have investigated *PRKCE* and its transcripts, suggesting the need for further research. Our analysis also revealed that mutations near the genes *SPBTN4* (HR = 1.7, $p = 3.6 \times 10^{-3}$) and *C3orf20* (HR = 0.65, $p = 1.5 \times 10^{-2}$) are associated with survival and with changes in expression. Although both genes are expressed in other cancers, their connection with GBM has not been studied (Figure 4D, E). .

To demonstrate the framework's integrative power, we highlight a G>C somatic variant within a predicted NR2F2 motif (chr1:153923729; Figure 4A). The mutation disrupts the canonical GGACC/ACCTC pattern (JASPAR, $p = 1 \times 10^{-3}$), reduces model attention at the site, and lowers the predicted NR2F2 binding probability. This predicted functional disruption was strongly correlated with clinical outcomes; Kaplan–Meier analysis revealed that carriers of the mutant allele had significantly shorter survival (log-rank p.adj = 0.05). This TFBS lies less than 5 kb upstream of *MIDN*, a neurodevelopmental gene[71], suggesting a potential GBM-relevant regulatory mechanism[72] (Figure 4A).

We also analyzed variants in splice sites for their association with survival. Within splice sites, a mutation in the acceptor site of PARPBP[73] (a PARP1-binding protein) was linked to improved survival (HR = 0.54, p = 0.02). This finding is supported by isoform expression analysis, which revealed an increase in an isoform that excludes the downstream exon (Figure 4C), which is consistent with the predicted acceptor site disruption[74]. In contrast, a mutation in the donor site of MAPT (Tau) was associated with worse survival (HR = 1.7, p = 0.0077)[75, 76]. This finding also aligned with the isoform expression data, where an increase in an isoform that skips the first exon was observed in the mutants, as predicted by our model. A comprehensive list of all survival-associated variants and regions is available in Supplementary notes S2 and S3.

***DeepVRegulome Identifies Biologically Relevant TFBS Variants in GBM:*** Our analysis highlights several candidate variants predicted by DeepVRegulome in both TFBS and splice site regions, which possess substantial biological relevance, particularly in pathways critical to disease pathogenesis and progression. We assessed these candidate variants in TFBS regions via two complementary approaches: first, by identifying TFBS models with the highest evaluation accuracy; second, by focusing on TFBS models that predict the highest number of candidate variants. In both strategies, we prioritized models exhibiting both substantial variant counts and established roles in gene regulation and cancer biology.

For the first approach, the top 35 TFBS models on the basis of evaluation accuracy are shown in a sunburst plot for SNVs (Figure-5B) and indels (Figure-5C), where the right-hand horizontal x-axis shows the highest accuracy of the TFBS model, and it decreases anticlockwise. Analysis of these top-performing TFBS models revealed hundreds of candidate variants affecting

TFs with established roles in neurodevelopment, immune regulation, and cancer. Notable examples include EGR2[39], a critical factor in nerve cell differentiation; SPI1[77], which is crucial for immune system function; ZIC2[78], a factor implicated in glioblastoma tumor progression; and CTCFL[79], which is involved in cancer-associated chromatin remodeling.

In the second approach, we summarize the candidate variants and their associations from the top TFBS models on the basis of the number of predicted candidate variants. Among the 460 TFBS models, the top 50 models for predicting candidate variants are presented as stacked bar plots in Figure 5. The y-axis of the stacked bar plot represents the combined count of significant indel variants (Figure 5(E)) and SNV variants (Figure 5(D)), along with their associated DBSNP counts. The secondary y-axis, located on the right-hand side, depicts model accuracy as a line plot. For SNV variants, the top 5 TFBS models based on candidate variant counts are NONO[80] (87.40%), RAD51[59, 81, 82] (87.06%), ZBTB33[83] (86.78%), IKZF2 (86.79%), and SMARCE1[84] (86.71%), which identify 93, 78, 69, 50, and 48 significant SNV variants, respectively. These SNVs are associated with 87, 76, 67, 49, and 44 DBSNP entities. In terms of indel variants, the top 5 TFBS models based on candidate variant counts are ZNF384[85] (88.58%), FOXA1 (88.52%), ZBTB33 (86.78%), MTA2[86] (85.15%), and NR3C1 (85.91%), which identify 368, 201, 148, 134, and 132 significant indel variants, respectively. These indels are associated with 158, 120, 115, 57, and 99 DBSNP entries. ZBTB33 (also known as Kaiso)[83]—the top-ranked model by both SNV and indel counts—is a CpG-methylation–dependent repressor that links epigenetic marks to transcriptional control, notably regulating Wnt-signaling genes implicated in cancer.

Furthermore, our analysis identified transcription factors with significant, known roles in glioblastoma biology. STAT3[87], a key oncogene in glioblastoma, was found to have 66 significant indel variants in its binding sites. Disruptions in the binding sites of FOXA1[88] and CREB1[89], which are involved in tumor invasion and proliferation, are also numerous. RFX1[90], another immune-related factor, is essential for MHC class II gene regulation, with implications in immune-mediated cancers and potential relevance to glioblastoma through its regulatory influence on immune pathways. Together, these TFBS models represent critical factors in cancer progression, chromatin remodeling, and immune response modulation, providing insights into pathways essential for glioblastoma and other cancer studies. A comprehensive analysis of the top 462 TFBS models for each approach is provided in Supplementary **Table S1**.

# 5 Discussion

DeepVRegulome provides a unified deep learning framework for genome-wide analyses of short variants for prioritization and interpretation of functionally disruptive candidate variants in noncoding regulatory genomic regions. DNABERT fine-tuned models, trained on the human regulome consisting of target regions of 700 different TFs, histone proteins and RBPs and splice junction regions, integrate gene regulatory information with high predictive accuracy across diverse regulatory contexts (Figure 2). Unlike classical motif scanners, which rely on fixed position-weight matrices and early deep-learning models trained for a single regulatory class, DeepVRegulome deploys element-specific transformer models for splice sites, TFBSs and histone marks in a single framework, capturing longer-range sequence grammar than previous methods do[23, 91, 92]. Importantly, we showed that the framework is not a "black box"; its integrated interpretation module learns biologically meaningful features, such as the successful rediscovery of known TF motifs from the JASPAR database. When applied to 190 glioblastoma cancer genomes, the approach prioritized ~**59,000** high-impact variants, >**10,000** of which recur in ≥10% of patients and are associated with overall survival (Supplementary Table S4). These results demonstrate that modern genome foundation models can bridge the gap between variant discovery and biological interpretation in large, clinically heterogeneous complex disease cohorts.

The strengths of DeepVRegulome can be categorized into three key areas. First, its region-specific fine-tuning strategy yields superior performance on short regulatory windows at a fraction of the input context used by long-range genomic models, enabling precise localization of motif disruptions. Second, the attention-guided interpretability module recovers canonical motifs de novo (Fig. 3) and links variant-level attention loss to experimentally validated TF motifs, providing mechanistic insight beyond probabilistic scores. Third, the pipeline integrates variant recurrence, dbSNP overlap and Kaplan–Meier statistics, converting sequence-level disruptions into clinically actionable signatures. For example, a G→C mutation that abolishes an *NR2F2* motif upstream of *MIDN* stratifies patient survival (log-rank $p = 0.0036$; Figure-4A), whereas an acceptor-site SNV in *PARPBP* predicts improved outcome and suggests heightened sensitivity to *PARP* inhibition. These findings highlight how the framework can guide novel therapeutic hypothesis-driven targets directly from WGS data.

The SNP-SELEX benchmarking places DeepVRegulome's variant-effect predictions in context relative to genomic foundation models that process orders of magnitude more input sequence. On the 17 transcription factors scored by every method, DeepVRegulome was statistically indistinguishable from Enformer, Borzoi, and AlphaGenome despite using an input window of 301 to 512 bp, roughly 2,000 times smaller than AlphaGenome's 1 Mb context. On broader factor sets, a gradient emerged in which models with longer input context achieved modestly higher mean AUROC. This pattern is consistent with the observation that TF binding events and local chromatin accessibility can be predicted accurately from approximately 500 to 2,000 bp of flanking sequence, because the relevant signal (the binding motif and its immediate sequence context) is local; longer-range context becomes necessary only for predicting distal regulatory outputs such as RNA expression levels, which integrate promoter, enhancer, and intervening chromatin signals across hundreds of kilobases [Borzoi ref]. DeepVRegulome's input window falls squarely within the range shown to be sufficient for TF binding prediction, and the modest, non-significant improvement observed when extending from 301 bp to 512 bp ($p = 0.076$) reinforces this conclusion.

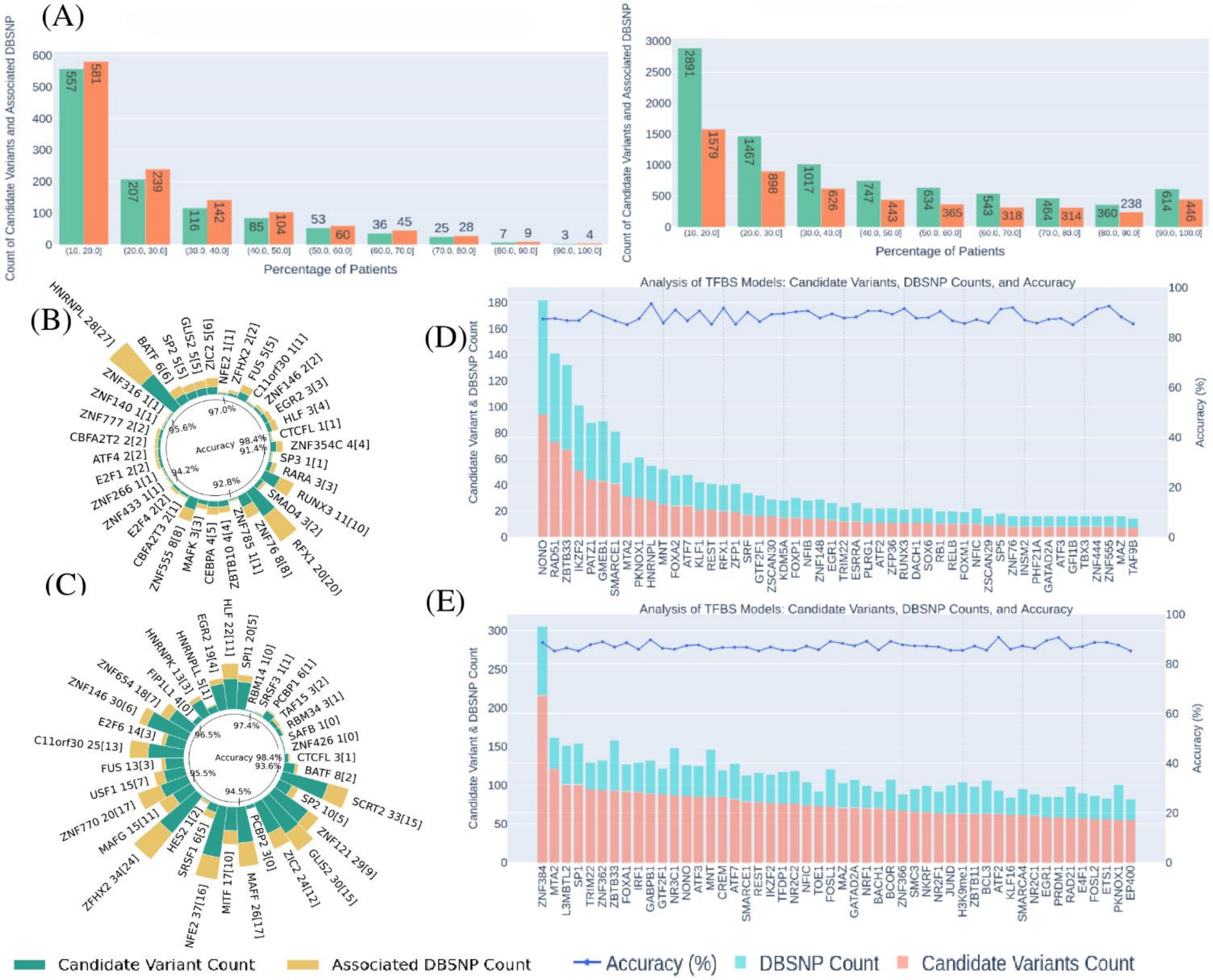

**Fig 5:** **(A)** Distribution of candidate and dbSNP-associated variants across GBM patients, stratified by the percentage of individuals affected. Left: CaVEMan SNVs; Right: Sanger Pindel indels. Candidate variants (green) and dbSNP variants (orange) are plotted per patient frequency bin, revealing that SNVs are more often annotated in dbSNP, while indels are predominantly novel. **(B, C)** Circular representation of the top 50 TFBS models based on model performance, sorted in decreasing order of evaluation accuracy. Each bar represents a TFBS model and is labeled as: *TFBS_model_name Candidate_Variant_Count [DBSNP_Count]*. Green bars indicate the number of candidate variants, yellow bars show associated dbSNP counts. Panel (B) shows CaVEMan SNVs, and panel (C) shows Sanger Pindel indels. **(D, E)** Comparison of the top 50 TFBS models based on the distribution of candidate variants, dbSNP counts, and model accuracy. Stacked bar plots show the number of candidate and dbSNP variants for each TFBS model (sorted in decreasing order of candidate variant count), while a blue line indicates model accuracy on a secondary y-axis. Panel (D) represents CaVEMan-based SNVs, and panel (E) represents Pindel-based indels.

The remaining performance gap between DeepVRegulome and AlphaGenome on broader factor sets reflects two independent factors that should be distinguished. First, input context length: AlphaGenome processes approximately 1 Mb of sequence, and its own 1 Mb configuration significantly outperforms its 16 kbp configuration on this dataset ($p = 0.003$), indicating that additional flanking sequence carries signal for a subset of factors. Second, training objective: AlphaGenome is trained to predict RNA-seq expression across many tissues and individuals in GTEx, which exposes the model to the effects of genetic variation on gene regulation through multi-genotype expression prediction, even though SNP-SELEX data were never part of AlphaGenome's training or evaluation corpus. DeepVRegulome's binary classification objective (bound vs. unbound on a single reference genome) provides no exposure to allelic variation during training.

This difference in inductive bias, rather than data leakage, likely accounts for much of the gap. Supporting this interpretation, a threshold analysis shows that AlphaGenome's advantage concentrates among factors with few positive examples and narrows from a mean AUROC difference of 0.067 at $n_positive \geq 5$ to 0.028 at $n_positive \geq 200$, where the difference is no longer significant ($p = 0.099$). Future work coupling DeepVRegulome's per-factor models with multi-task pretraining on genotype-expression data could combine the framework's factor-specific precision with the variant-sensitivity advantages of foundation-model training objectives.

A major finding of this study is the large "annotation gap" for functional noncoding variants. We found that the vast majority of the high-impact, recurrent variants identified by DeepVRegulome are not recorded in clinical databases such as ClinVar. The few that were found, such as in the TFBSs for *SMAD4* and *FOXM1*, these TFs have well-established roles in cancer progression. This highlights the vast, clinically uncharacterized landscape of potentially pathogenic noncoding variants and establishes DeepVRegulome as a powerful discovery tool for prioritizing these variants for future clinical annotation and research. Beyond oncology, DeepVRegulome is readily extensible to germline rare-disease sequencing, population-scale variant catalogs and multimodal datasets that include methylation or chromatin accessibility profiles. The open-source code, trained weights and interactive dashboard (data availability) lower the barrier for community adoption and for benchmarking against emerging foundation models. We anticipate that systematic application of DeepVRegulome across large consortia cohorts—such as PCAWG, TCGA-PanCancer and UK Biobank—will yield a comprehensive atlas of functional noncoding mutations, thereby accelerating variant prioritization for precision medicine.

Several limitations merit discussion. DeepVRegulome uses input windows of 301 to 512 bp, which, as the SNP-SELEX benchmarking confirms, capture the local sequence grammar sufficient for TF binding prediction but may miss distal enhancer-promoter interactions and long-range chromatin loops that influence gene regulation at greater distances. This local focus is also the likely reason for the framework's lower performance on most histone marks than on TFBSs; histone modifications are often governed by broader regional context or DNA shape rather than the specific local motifs our model excels at identifying. Extending to context-aware models such as DNABERT-2 or HyenaDNA[7,8]—and incorporating Hi-C-derived contact priors—should improve the detection of distal regulatory effects. Our study is confined to glioblastoma; although the model architecture is agnostic to tumor type, additional cancer cohorts will be needed to assess generalizability and to refine recurrence thresholds. Finally, our functional validation is confined to in silico motif disruption, RNA-seq isoform shifts, and survival correlations; establishing causality will require orthogonal wet-lab assays—such as allelic luciferase or minigene reporters, CRISPR base editing of glioblastoma lines, and single-cell transcriptomics—to confirm that the identified variants directly perturb regulatory activity.

In summary, DeepVRegulome couples high-fidelity regulatory modeling with transparent interpretation and clinical association, establishing an end–end path from raw whole-genome sequencing to biologically and clinically meaningful variant sets. By illuminating the regulatory genome at single-nucleotide resolution, the framework advances our ability to decipher the noncoding mutational landscape and highlights new avenues for translational genomics.

# 6 Code and data availability

The DeepVRegulome source code is available at https://github.com/DavuluriLab/DeepVRegulome. The fine-tuned model checkpoints and intermediate files will be deposited in Zenodo and released under a CC-BY-NC license upon article acceptance; they are available from the corresponding author upon reasonable request during peer review. An interactive web application for exploring candidate variants, associated survival data, and gene-level information is accessible at https://deepvregulome.streamlit.app. The framework is also available as an installable Python package (pip install deepvregulome) and the fine-tuned model weights are hosted on the Hugging Face model hub (https://huggingface.co/duttaprat/DeepVRegulome).

# Conclusion

In this study, we present **DeepVRegulome**, a scalable and interpretable deep learning framework designed to accurately prioritize clinically relevant somatic mutations in noncoding regulatory regions of the human genome. By fine-tuning DNABERT on splice sites from GENCODE and transcription factor-binding site datasets from ENCODE, we achieve highly accurate functional predictions across diverse regulatory elements. Benchmarking against an independent experimental assay of allele-specific transcription factor binding (SNP-SELEX) confirmed that these predictions track real variant effects on binding, with performance matching or approaching that of substantially larger context-based models despite using a much smaller input window. The integrated visualization module reveals biologically meaningful sequence motifs and attention patterns, facilitating direct interpretation in a clinical genomics context. Applying DeepVRegulome to glioblastoma whole-genome sequencing data revealed recurrently mutated regulatory regions significantly associated with patient survival, underscoring its potential as novel prognostic markers. Importantly, DeepVRegulome's modular and data-agnostic design allows straightforward adaptation to other cancer types and complex diseases. By bridging genome-scale variant discovery with actionable clinical insights, DeepVRegulome significantly advances the clinical interpretation of noncoding genomic variation, paving the way toward precision oncology and personalized genomic medicine.